\documentclass[twocolumn]{aastex63}

\submitjournal{ApJL}

\shorttitle{The curious case of PHL 293B}
\shortauthors{Burke et al.}
\graphicspath{{./}{figures/}}

\begin{document}

\title{The Curious Case of PHL 293B: A Long-Lived Transient in a Metal-Poor Blue Compact Dwarf Galaxy}

\correspondingauthor{Colin J. Burke}
\email{colinjb2@illinois.edu}

\suppressAffiliations
\author[0000-0001-9947-6911]{Colin~J.~Burke}
\affiliation{Department of Astronomy, University of Illinois at Urbana-Champaign, 1002 W. Green Street, Urbana, IL 61801, USA}
\affiliation{National Center for Supercomputing Applications, 
1205 West Clark Street, Urbana, IL 61801, USA}

\author[0000-0003-4703-7276]{Vivienne~F.~Baldassare}
\altaffiliation{Einstein Fellow}
\affiliation{Department of Astronomy, Yale University,
52 Hillhouse Avenue, New Haven, CT 06511, USA}

\author[0000-0003-0049-5210]{Xin~Liu}
\affiliation{Department of Astronomy, University of Illinois at Urbana-Champaign, 1002 W. Green Street, Urbana, IL 61801, USA}
\affiliation{National Center for Supercomputing Applications, 
1205 West Clark Street, Urbana, IL 61801, USA}

\author{Ryan~J.~Foley}
\affiliation{Department of Astronomy and Astrophysics, University of California, Santa Cruz, CA 95064, USA}

\author[0000-0003-1659-7035]{Yue~Shen}
\altaffiliation{Alfred P. Sloan Fellow}
\affiliation{Department of Astronomy, University of Illinois at Urbana-Champaign, 1002 W. Green Street, Urbana, IL 61801, USA}
\affiliation{National Center for Supercomputing Applications, 
1205 West Clark Street, Urbana, IL 61801, USA}

\author[0000-0002-6011-0530]{Antonella~Palmese}
\affiliation{Fermi National Accelerator Laboratory, P. O. Box 500, Batavia, IL 60510, USA}
\affiliation{Kavli Institute for Cosmological Physics, University of Chicago, 5640 South Ellis Avenue, Chicago, IL 60637, USA}

\author[0000-0001-8416-7059]{Hengxiao~Guo}
\affiliation{Department of Astronomy, University of Illinois at Urbana-Champaign, 1002 W. Green Street, Urbana, IL 61801, USA}
\affiliation{National Center for Supercomputing Applications, 
1205 West Clark Street, Urbana, IL 61801, USA}

\author{Kenneth~Herner}
\affiliation{Fermi National Accelerator Laboratory, P. O. Box 500, Batavia, IL 60510, USA}


\author{Tim~M.~C.~Abbott}
\affiliation{Cerro Tololo Inter-American Observatory, NSF’s National Optical-Infrared Astronomy Research Laboratory, Casilla 603, La Serena, Chile}
\author{Michel~Aguena}
\affiliation{Departamento de F\'isica Matem\'atica, Instituto de F\'isica, Universidade de S\~ao Paulo, CP 66318, S\~ao Paulo, SP, 05314-970, Brazil}
\affiliation{Laborat\'orio Interinstitucional de e-Astronomia - LIneA, Rua Gal. Jos\'e Cristino 77, Rio de Janeiro, RJ - 20921-400, Brazil}
\author{Sahar~Allam}
\affiliation{Fermi National Accelerator Laboratory, P. O. Box 500, Batavia, IL 60510, USA}
\author{Santiago~Avila}
\affiliation{Instituto de Fisica Teorica UAM/CSIC, Universidad Autonoma de Madrid, 28049 Madrid, Spain}
\author{Emmanuel~Bertin}
\affiliation{CNRS, UMR 7095, Institut d'Astrophysique de Paris, F-75014, Paris, France}
\affiliation{Sorbonne Universit\'es, UPMC Univ Paris 06, UMR 7095, Institut d'Astrophysique de Paris, F-75014, Paris, France}
\author{David~Brooks}
\affiliation{Department of Physics \& Astronomy, University College London, Gower Street, London, WC1E 6BT, UK}
\author{Aurelio~Carnero~Rosell}
\affiliation{Centro de Investigaciones Energ\'eticas, Medioambientales y Tecnol\'ogicas (CIEMAT), Madrid, Spain}
\author{Matias~Carrasco~Kind}
\affiliation{Department of Astronomy, University of Illinois at Urbana-Champaign, 1002 W. Green Street, Urbana, IL 61801, USA}
\affiliation{National Center for Supercomputing Applications, 1205 West Clark St., Urbana, IL 61801, USA}
\author{Jorge~Carretero}
\affiliation{Institut de F\'{\i}sica d'Altes Energies (IFAE), The Barcelona Institute of Science and Technology, Campus UAB, 08193 Bellaterra (Barcelona) Spain}
\author{Luiz~N.~da Costa}
\affiliation{Laborat\'orio Interinstitucional de e-Astronomia - LIneA, Rua Gal. Jos\'e Cristino 77, Rio de Janeiro, RJ - 20921-400, Brazil}
\affiliation{Observat\'orio Nacional, Rua Gal. Jos\'e Cristino 77, Rio de Janeiro, RJ - 20921-400, Brazil}
\author{Juan~De~Vicente}
\affiliation{Centro de Investigaciones Energ\'eticas, Medioambientales y Tecnol\'ogicas (CIEMAT), Madrid, Spain}
\author{Shantanu~Desai}
\affiliation{Department of Physics, IIT Hyderabad, Kandi, Telangana 502285, India}
\author{Peter~Doel}
\affiliation{Department of Physics \& Astronomy, University College London, Gower Street, London, WC1E 6BT, UK}
\author{Tim~F.~Eifler}
\affiliation{Department of Astronomy/Steward Observatory, University of Arizona, 933 North Cherry Avenue, Tucson, AZ 85721-0065, USA}
\affiliation{Jet Propulsion Laboratory, California Institute of Technology, 4800 Oak Grove Dr., Pasadena, CA 91109, USA}
\author{Spencer~Everett}
\affiliation{Santa Cruz Institute for Particle Physics, Santa Cruz, CA 95064, USA}
\author{Josh~Frieman}
\affiliation{Fermi National Accelerator Laboratory, P. O. Box 500, Batavia, IL 60510, USA}
\affiliation{Kavli Institute for Cosmological Physics, University of Chicago, Chicago, IL 60637, USA}
\author{Juan~Garc\'ia-Bellido}
\affiliation{Instituto de Fisica Teorica UAM/CSIC, Universidad Autonoma de Madrid, 28049 Madrid, Spain}
\author{Enrique~Gaztanaga}
\affiliation{Institut d'Estudis Espacials de Catalunya (IEEC), 08034 Barcelona, Spain}
\affiliation{Institute of Space Sciences (ICE, CSIC),  Campus UAB, Carrer de Can Magrans, s/n,  08193 Barcelona, Spain}
\author{Daniel~Gruen}
\affiliation{Department of Physics, Stanford University, 382 Via Pueblo Mall, Stanford, CA 94305, USA}
\affiliation{Kavli Institute for Particle Astrophysics \& Cosmology, P. O. Box 2450, Stanford University, Stanford, CA 94305, USA}
\affiliation{SLAC National Accelerator Laboratory, Menlo Park, CA 94025, USA}
\author{Robert~A.~Gruendl}
\affiliation{Department of Astronomy, University of Illinois at Urbana-Champaign, 1002 W. Green Street, Urbana, IL 61801, USA}
\affiliation{National Center for Supercomputing Applications, 1205 West Clark St., Urbana, IL 61801, USA}
\author{Julia~Gschwend}
\affiliation{Laborat\'orio Interinstitucional de e-Astronomia - LIneA, Rua Gal. Jos\'e Cristino 77, Rio de Janeiro, RJ - 20921-400, Brazil}
\affiliation{Observat\'orio Nacional, Rua Gal. Jos\'e Cristino 77, Rio de Janeiro, RJ - 20921-400, Brazil}
\author{Gaston~Gutierrez}
\affiliation{Fermi National Accelerator Laboratory, P. O. Box 500, Batavia, IL 60510, USA}
\author{Devon~L.~Hollowood}
\affiliation{Santa Cruz Institute for Particle Physics, Santa Cruz, CA 95064, USA}
\author{Klaus~Honscheid}
\affiliation{Center for Cosmology and Astro-Particle Physics, The Ohio State University, Columbus, OH 43210, USA}
\affiliation{Department of Physics, The Ohio State University, Columbus, OH 43210, USA}
\author{David~J.~James}
\affiliation{Center for Astrophysics $\vert$ Harvard \& Smithsonian, 60 Garden Street, Cambridge, MA 02138, USA}
\author{Elisabeth~Krause}
\affiliation{Department of Astronomy/Steward Observatory, University of Arizona, 933 North Cherry Avenue, Tucson, AZ 85721-0065, USA}
\author{Kyler~Kuehn}
\affiliation{Australian Astronomical Optics, Macquarie University, North Ryde, NSW 2113, Australia}
\affiliation{Lowell Observatory, 1400 Mars Hill Rd, Flagstaff, AZ 86001, USA}
\author{Marcio~A.~G.~Maia}
\affiliation{Laborat\'orio Interinstitucional de e-Astronomia - LIneA, Rua Gal. Jos\'e Cristino 77, Rio de Janeiro, RJ - 20921-400, Brazil}
\affiliation{Observat\'orio Nacional, Rua Gal. Jos\'e Cristino 77, Rio de Janeiro, RJ - 20921-400, Brazil}
\author{Felipe~Menanteau}
\affiliation{Department of Astronomy, University of Illinois at Urbana-Champaign, 1002 W. Green Street, Urbana, IL 61801, USA}
\affiliation{National Center for Supercomputing Applications, 1205 West Clark St., Urbana, IL 61801, USA}
\author{Ramon~Miquel}
\affiliation{Instituci\'o Catalana de Recerca i Estudis Avan\c{c}ats, E-08010 Barcelona, Spain}
\affiliation{Institut de F\'{\i}sica d'Altes Energies (IFAE), The Barcelona Institute of Science and Technology, Campus UAB, 08193 Bellaterra (Barcelona) Spain}
\author{Francisco~Paz-Chinch\'{o}n}
\affiliation{Institute of Astronomy, University of Cambridge, Madingley Road, Cambridge CB3 0HA, UK}
\affiliation{National Center for Supercomputing Applications, 1205 West Clark St., Urbana, IL 61801, USA}
\author{Andrés~A.~Plazas}
\affiliation{Department of Astrophysical Sciences, Princeton University, Peyton Hall, Princeton, NJ 08544, USA}
\author{Eusebio~Sanchez}
\affiliation{Centro de Investigaciones Energ\'eticas, Medioambientales y Tecnol\'ogicas (CIEMAT), Madrid, Spain}
\author{Basilio~Santiago}
\affiliation{Instituto de F\'\i sica, UFRGS, Caixa Postal 15051, Porto Alegre, RS - 91501-970, Brazil}
\affiliation{Laborat\'orio Interinstitucional de e-Astronomia - LIneA, Rua Gal. Jos\'e Cristino 77, Rio de Janeiro, RJ - 20921-400, Brazil}
\author{Vic~Scarpine}
\affiliation{Fermi National Accelerator Laboratory, P. O. Box 500, Batavia, IL 60510, USA}
\author{Santiago~Serrano}
\affiliation{Institut d'Estudis Espacials de Catalunya (IEEC), 08034 Barcelona, Spain}
\affiliation{Institute of Space Sciences (ICE, CSIC),  Campus UAB, Carrer de Can Magrans, s/n,  08193 Barcelona, Spain}
\author{Ignacio~Sevilla-Noarbe}
\affiliation{Centro de Investigaciones Energ\'eticas, Medioambientales y Tecnol\'ogicas (CIEMAT), Madrid, Spain}
\author{Mathew~Smith}
\affiliation{School of Physics and Astronomy, University of Southampton,  Southampton, SO17 1BJ, UK}
\author{Marcelle~Soares-Santos}
\affiliation{Brandeis University, Physics Department, 415 South Street, Waltham MA 02453}
\author{Eric~Suchyta}
\affiliation{Computer Science and Mathematics Division, Oak Ridge National Laboratory, Oak Ridge, TN 37831}
\author{Molly~E.~C.~Swanson}
\affiliation{National Center for Supercomputing Applications, 1205 West Clark St., Urbana, IL 61801, USA}
\author{Gregory~Tarle}
\affiliation{Department of Physics, University of Michigan, Ann Arbor, MI 48109, USA}
\author{Douglas~L.~Tucker}
\affiliation{Fermi National Accelerator Laboratory, P. O. Box 500, Batavia, IL 60510, USA}
\author{Tamas~Norbert~Varga}
\affiliation{Max Planck Institute for Extraterrestrial Physics, Giessenbachstrasse, 85748 Garching, Germany}
\affiliation{Universit\"ats-Sternwarte, Fakult\"at f\"ur Physik, Ludwig-Maximilians Universit\"at M\"unchen, Scheinerstr. 1, 81679 M\"unchen, Germany}
\author{Alistair~R.~Walker}
\affiliation{Cerro Tololo Inter-American Observatory, NSF’s National Optical-Infrared Astronomy Research Laboratory, Casilla 603, La Serena, Chile}

\collaboration{53}{(DES Collaboration)}



\begin{abstract}

We report on small-amplitude optical variability and recent dissipation of the unusually persistent broad emission lines in the blue compact dwarf galaxy PHL 293B. The galaxy's unusual spectral features (P Cygni-like profiles with $\sim$800 km s$^{-1}$ blueshifted absorption lines) have resulted in conflicting interpretations of the nature of this source in the literature. However, analysis of new Gemini spectroscopy reveals the broad emission has begun to fade after being persistent for over a decade prior. Precise difference imaging light curves constructed with the Sloan Digital Sky Survey and the Dark Energy Survey reveal small-amplitude optical variability of $\sim$0.1 mag in the \emph{g} band offset by $100\pm21$ pc from the brightest pixel of the host. The light curve is well-described by an active galactic nuclei (AGN)-like damped random walk process. However, we conclude that the origin of the optical variability and spectral features of PHL 293B is due to a long-lived stellar transient, likely a Type IIn supernova or non-terminal outburst, mimicking long-term AGN-like variability. This work highlights the challenges of discriminating between scenarios in such extreme environments, relevant to searches for AGNs in dwarf galaxies. This is the second long-lived transient discovered in a blue compact dwarf, after SDSS1133. Our result implies such long-lived stellar transients may be more common in metal-deficient galaxies. Systematic searches for low-level variability in dwarf galaxies will be possible with the upcoming Legacy Survey of Space and Time at Vera C. Rubin Observatory.

\end{abstract}


\reportnum{DES-2020-0524}
\reportnum{FERMILAB-PUB-20-068-AE}


\section{Introduction} \label{sec:intro}

Blue compact dwarf (BCD) galaxies \citep{Thuan1981}, particularly metal-poor ones, are important laboratories for studying galaxies in their earliest stages of evolution. They may be undergoing their first round of star formation, containing massive O and B stars responsible for their blue colors. Therefore, BCD galaxies may act as analogues to primordial high redshift galaxies, offering unique opportunities to study intense star formation and low-metallicity environments.

PHL 293B\footnote{Also known as the Kinman dwarf or SDSS J223036.79-000636.9.} is a metal-poor ($12+\log{{\rm O}/{\rm H}}= 7.71 \pm 0.02$; \citealt{Izotov2011}) BCD emission-line galaxy at $z=0.00517 \pm 0.00001$ (NED)\footnote{For consistency with \citet{Terlevich2014}, a distance of 23.1 Mpc is adopted throughout (corrected for the Virgo cluster + Great Attractor + Shapley) using $H_0=73.0$ km s$^{-1}$ Mpc$^{-1}$, $\Omega_m=0.27$, and , $\Omega_\Lambda=0.73$. See \url{http://ned.ipac.caltech.edu/byname?objname=kinman\%20dwarf&hconst=73.0&omegam=0.27&omegav=0.73&wmap=1&corr_z=4}.}. PHL 293B has a stellar mass of $\sim2\times10^{7}~M_{\odot}$ and an HI gas fraction of 0.75 \citep{2006ApJ...653..240G}. The source previously exhibited striking P Cygni-like broad emission in Balmer series lines \citep{Izotov2009,Izotov2011}. The narrow absorption lines were blueshifted by $\sim$800 km s$^{-1}$ and the broad H$\alpha$ emission had a FWHM of about 1500 km s$^{-1}$. The spectral features (including broad lines) persisted for over a decade until only recently. The origin of these features has been a source of speculation, with conflicting interpretations in the literature. A luminous blue variable (LBV) star outburst \citep{Izotov2009,Izotov2011,Allan2020}, an expanding supershell or stationary superwind driven by a young stellar wind \citep{Terlevich2014}, and a strongly-radiative stationary cooling wind driven by old supernova remnants \citep{Tenorio-Tagle2015} have been proposed.

Here we report on variability measured in the optical light curves using difference imaging on Sloan Digital Sky Survey (SDSS) and years 1 -- 6 (Y6) Dark Energy Survey \citep[DES;][]{Flaugher2015,DES2016,DW2018} images. The previously-undetected variability has an amplitude of 0.12 mag in the $\emph{g}$ band between 1998 and 2018. This observation rules out stationary winds and other non-transient interpretations for the mechanism of the broad emission lines in PHL 293B. The optical variability is well-described by a damped random walk process typical of active galactic nuclei (AGN) \citep{2009ApJ...698..895K, 2010ApJ...721.1014M}. However, analysis of recently-obtained Gemini \mbox{GMOS-N} spectroscopy reveals the broad emission component has now begun to subside. A changing-look AGN scenario is unlikely given the lack of X-ray emission and high-ionization lines, as discussed in \S\ref{sec:discussion}. Most likely, the variability is due to a transient event mimicking AGN-like variability.

In light of our new observations, we investigate several scenarios which could explain the features of PHL 293B. We conclude that the source of the variability and spectral features of PHL 293B is most likely due to a long-lived Type IIn supernova (SN IIn)-like transient. In particular, we speculate that we could be observing a similar event to the peculiar long-lived transient SDSS J113323.97+550415.8 (SDSS1133) in the BCD galaxy Mrk 177 discovered by \cite{Koss2014}. In this scenario, an LBV progenitor explodes as a SN IIn-like event, followed by a slowly flattening light curve and unusually persistent broad emission line features.

This Letter is organized as follows. In \S\ref{sec:newobs}, we present combined SDSS and DES light curves which show a slowly fading transient, along with new Gemini GMOS-N spectroscopy which shows the broad emission has faded recently. In \S\ref{sec:obs}, we summarize the existing observations of PHL 293B in the literature. In \S\ref{sec:discussion}, we review the interpretations of the nature of PHL 293B in the literature and attempt to reconcile its newly-observed photometric and spectroscopic variability with the previous work. In \S\ref{sec:conclusion}, we summarize our findings and conclude that, contrary to previous explanations, PHL 293B is most likely a long-lived SN IIn-like event.

\section{New Observations} \label{sec:newobs}

\subsection{SDSS+DES Light Curve} \label{sec:lc}

\begin{figure*}
\gridline{\fig{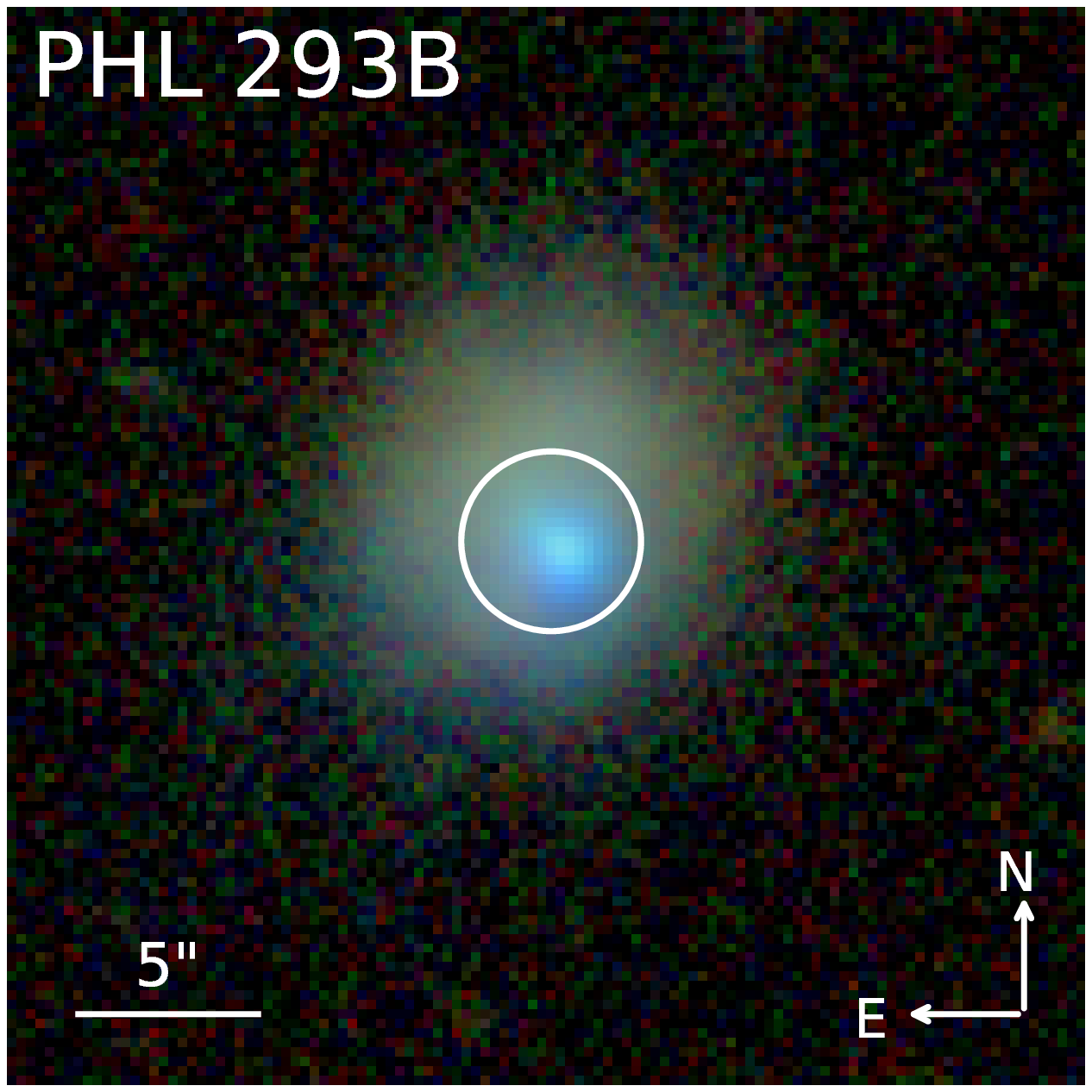}{0.3\textwidth}{(a)}
          \fig{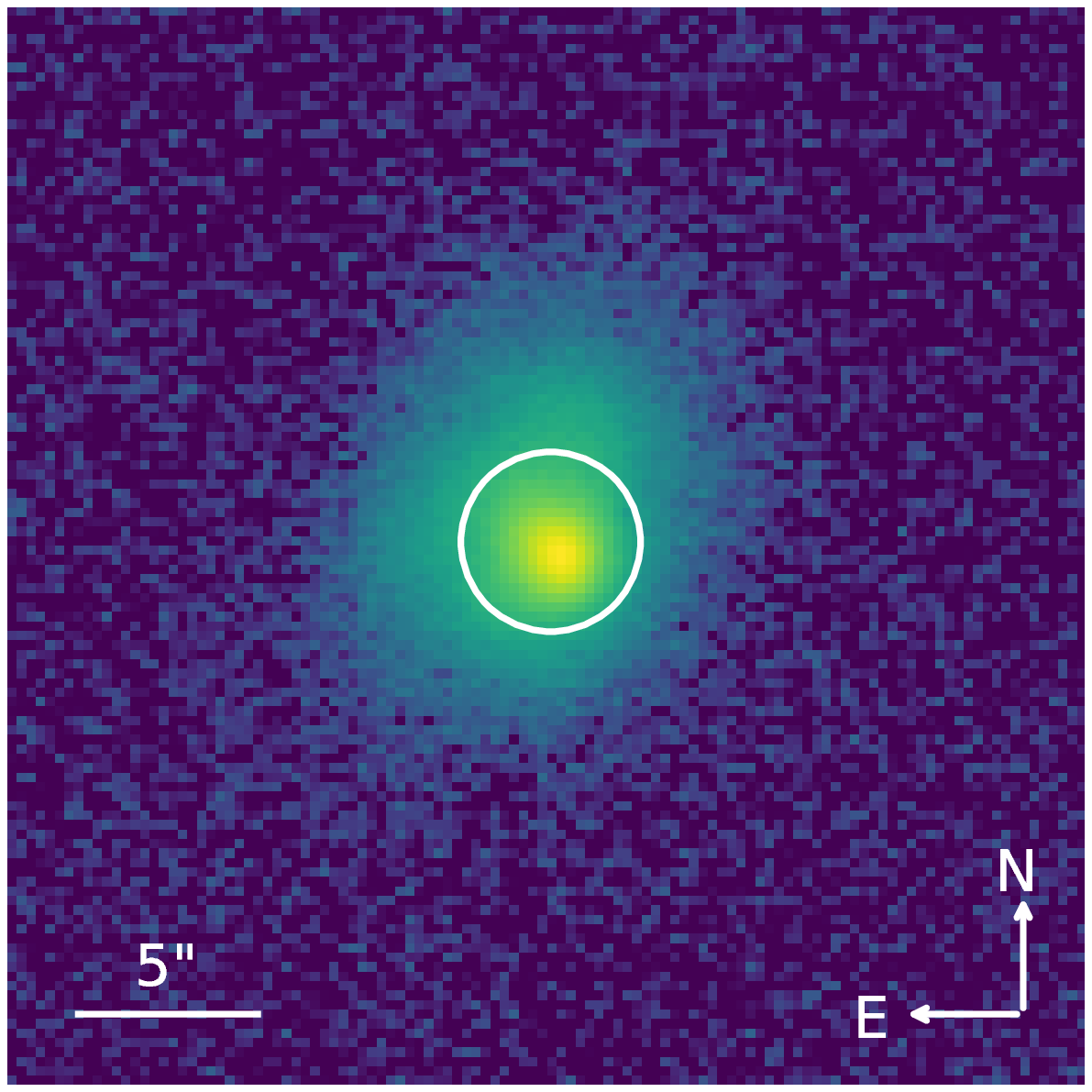}{0.3\textwidth}{(b)}
          \fig{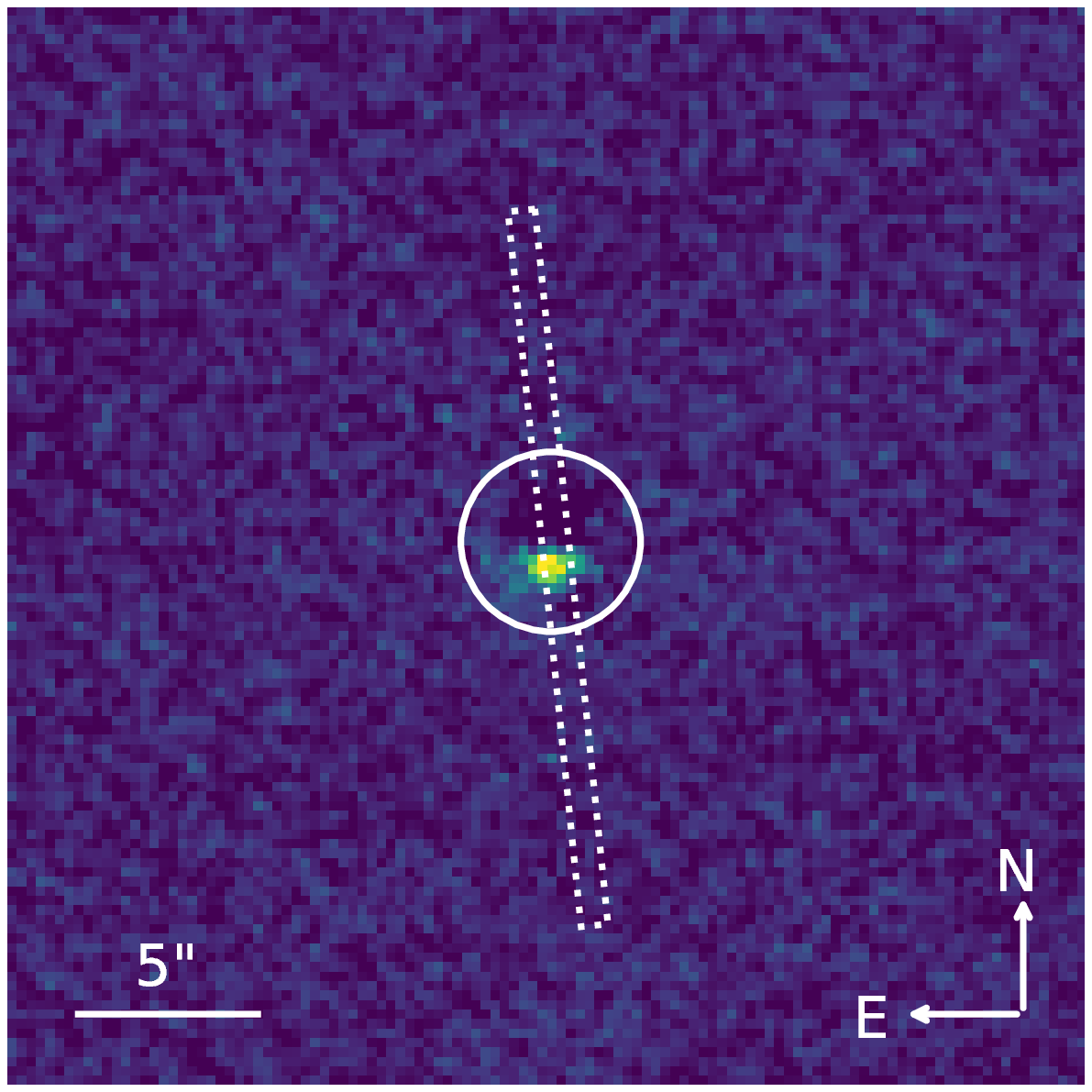}{0.3\textwidth}{(c)}
          }
\caption{Difference imaging analysis of PHL 293B photometry with SDSS and DES. The top row shows (a) the DES \emph{gri} color composite Y6 coadd, (b) the DES \emph{g} band template image, and (c) the DES \emph{g} band coadd of the difference images. The circles enclose the 2.5$^{\prime\prime}$ radius target aperture. The GMOS slit configuration is also shown in panel (c).  The difference images indicate a single variable point source offset by $100\pm21$ pc. \label{fig:0}}
\gridline{\fig{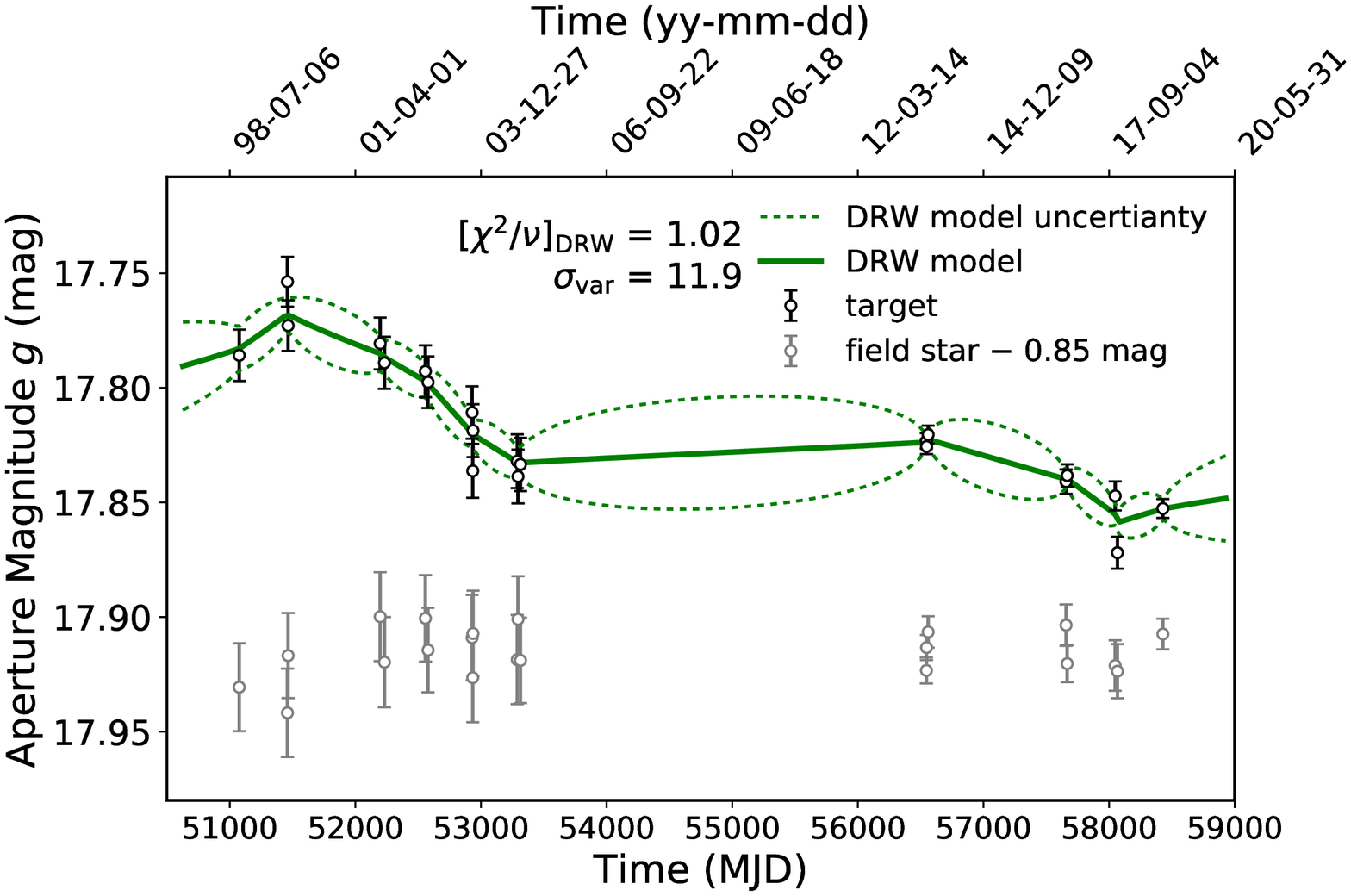}{0.5\textwidth}{(a)}
          \fig{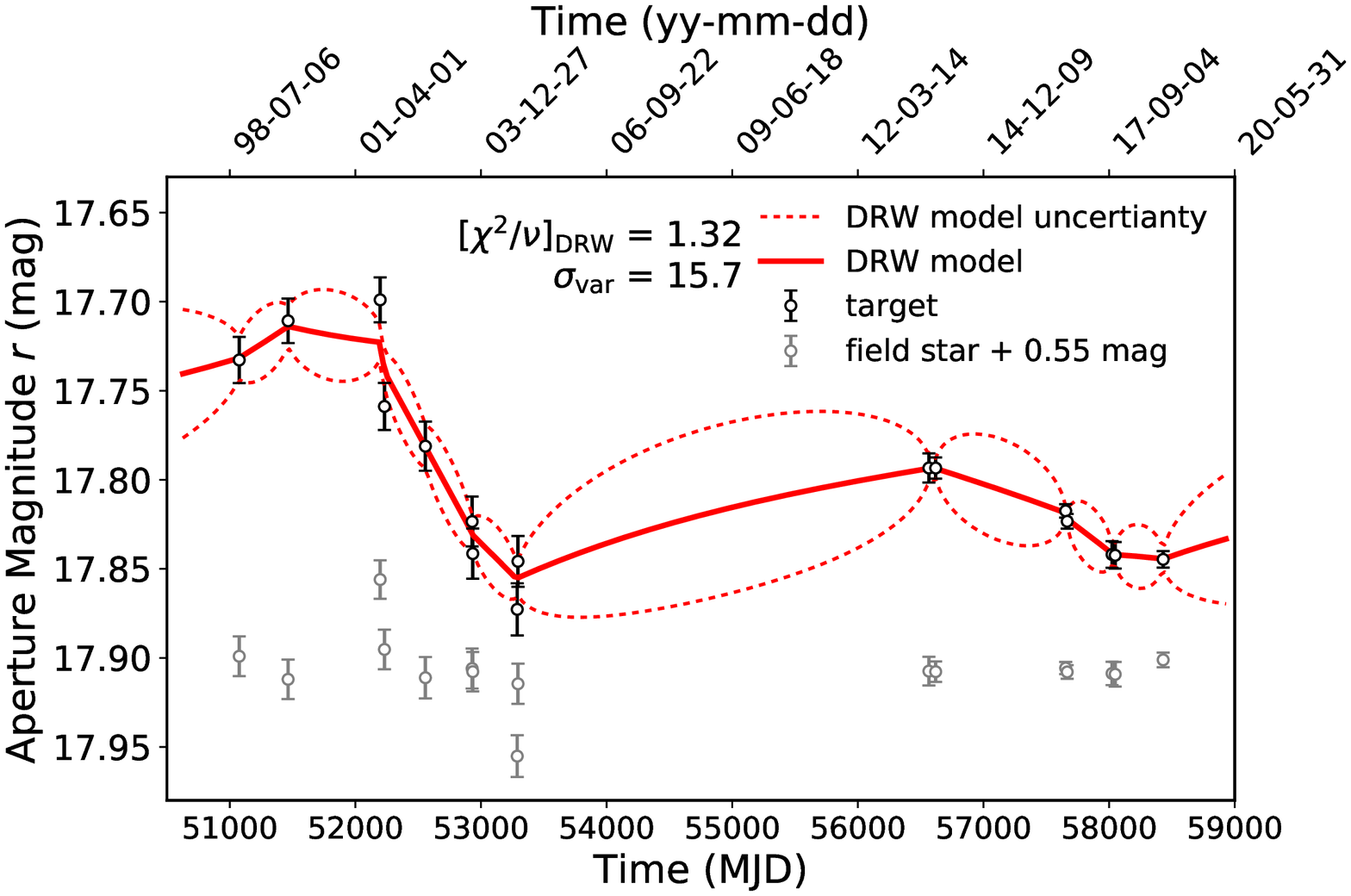}{0.5\textwidth}{(b)}
}
\caption{The SDSS and DES combined difference imaging light curves of PHL 293B (target) are shown in \emph{g} (a) and \emph{r} (b) bands. A nearby field star is shown for comparison. The best-fit damped random walk (DRW) model (solid lines) and model uncertainty (dashed lines) are shown. The light curves are constructed from SDSS and DES imaging between 1998 and 2018. \label{fig:1}}
\end{figure*}

The optical variability in PHL 298B was discovered independently in searches for AGN in dwarf galaxies in SDSS \citep{Baldassare2018} and in DES (Burke et al. in preparation). Here, we combine the SDSS and DES light curves for a total baseline spanning two decades (1998--2018). We perform standard difference image analysis (DIA) to isolate the variable point-source flux from seeing variations between epochs.

We summarize the analysis and features of the light curve here; see Appx.~\ref{apdx} for details. The DES coadd, template image, difference image coadd, and the DIA multi-band SDSS+DES light curves are shown in Fig.~\ref{fig:0}. The difference image flux is consistent with an unresolved point source offset by $3.5\pm0.7$ pixels or $100\pm21$ pc from the brightest pixel of the host. The SDSS photometry was taken between MJD 51075 and 53314 (between 1998 and 2005). The DES photometry was taken between MJD 56545 and 58428 (between 2013 and 2018). The amplitude is 0.12 mag (\emph{g} band) which corresponds to a $\sim10$ percent variation in luminosity over 20 years.

We originally selected PHL 293B as having variability consistent with an AGN. To test this in more detail, we fit the light curve to a damped random walk (DRW) model (generally a good empirical descriptor of AGN variability on days to years timescales; \citealt{2009ApJ...698..895K, 2010ApJ...721.1014M}). To assess the fit, we calculate the reduced $\chi^2$ of the DRW model $\left[\chi^2/\nu\right]_{\rm DRW}$. We also calculate the significance that the source is variable $\sigma_{\rm var}$ in units of $\sigma$ from a $\chi^2$ test given the photometry and uncertainties (see Appx.~\ref{apdx}). We find the $\emph{g}$-band variability is significant at the $11\sigma$ level, and the DRW model is a good fit to the data. However, the overall trend is a slow fading at a rate of $\sim$0.005 mag year$^{-1}$ in the \emph{g} band. Along with the evidence presented in \S\ref{sec:discussion}, we conclude that we are instead witnessing a long-lived transient mimicking AGN-like variability.

The nearby field star J223033.18-000633.7 is used for comparison and to correct for any zeropoint difference between the SDSS and DES photometric systems. The field star is not significantly variable (\emph{g}-band $\sigma_{\rm var}=0.16$) The variability of PHL 293B is marginally significant in the DES data and certainly present in SDSS. Inspection of the difference images by eye did not reveal any artifacts that would indicate a bad image subtraction. Furthermore, the same variable trend is present in all photometric bands, indicating it is not due to a systematic (Fig.~\ref{fig:1}).

\subsection{Gemini Spectroscopy} \label{sec:spec}

\begin{figure*}
\gridline{\fig{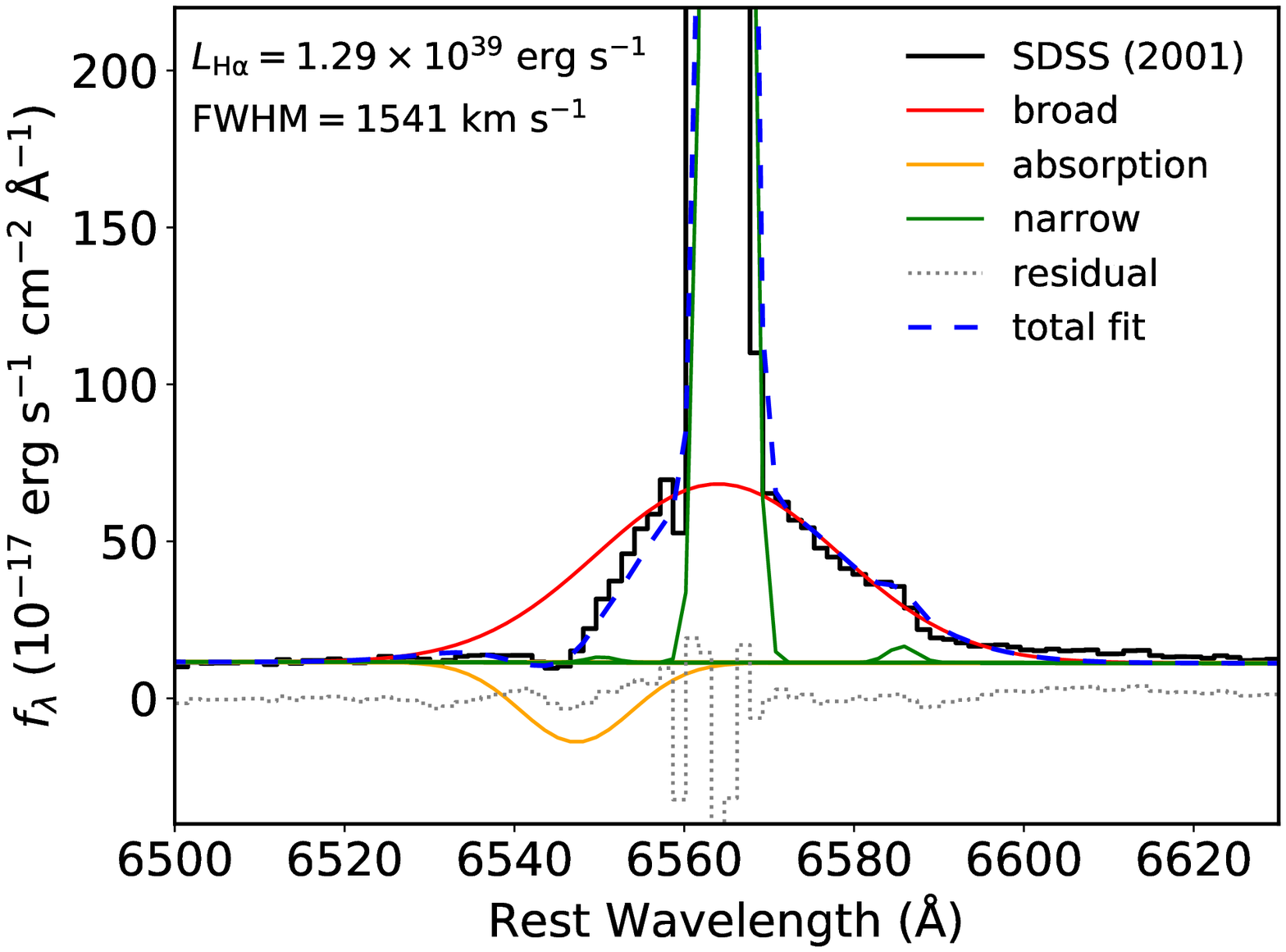}{0.5\textwidth}{(a)}
          \fig{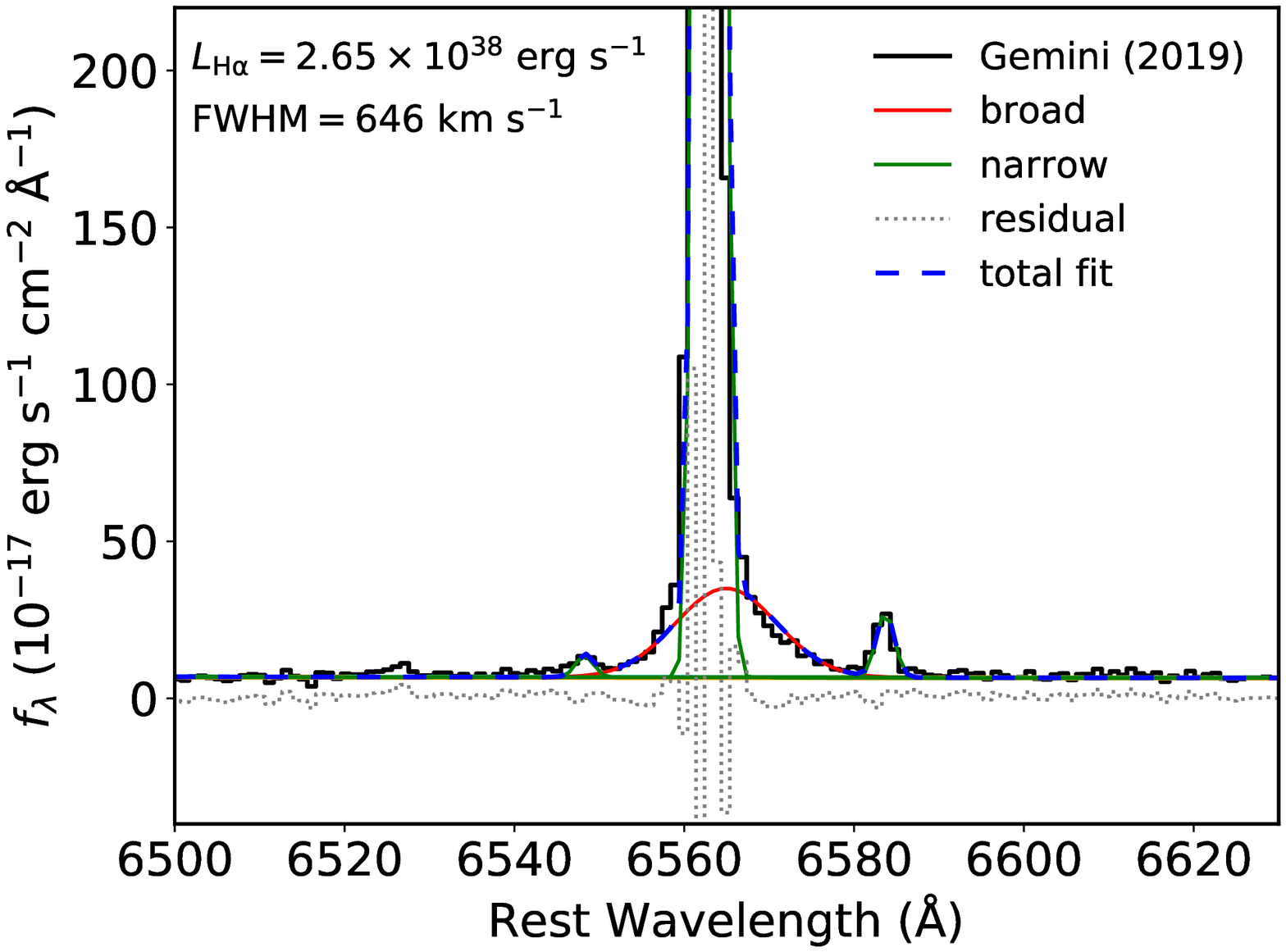}{0.5\textwidth}{(b)}
}
\gridline{\fig{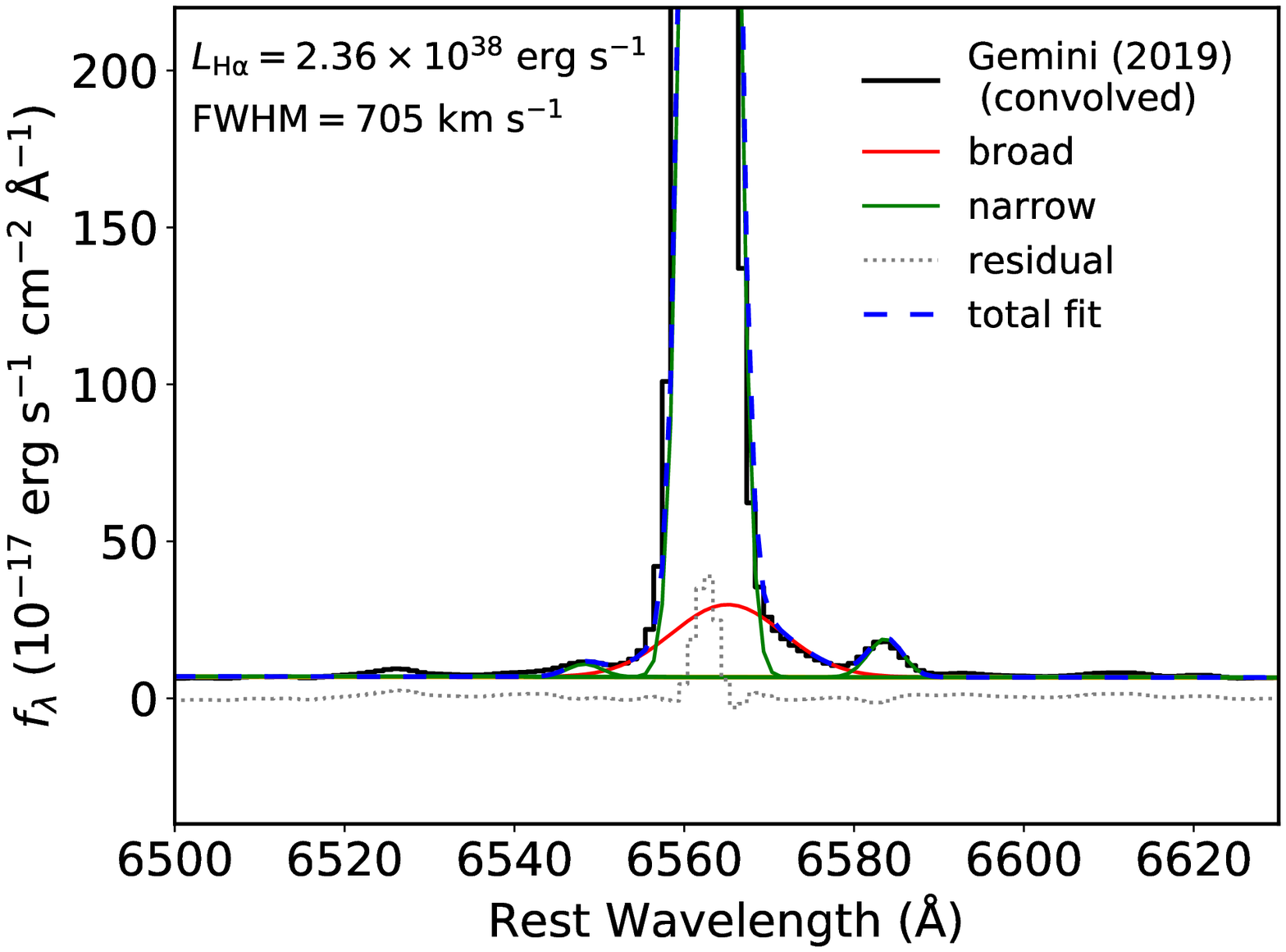}{0.5\textwidth}{(c)}
          }
\caption{H$\alpha$-[N~\textsc{ii}] complex from the SDSS (a) and Gemini (b) spectrum of PHL 293B. We convolved the Gemini spectrum with a Gaussian of width $\rm{FWHM}=1541$ km s$^{-1}$ to match the GMOS spectral resolution to SDSS. This is shown in panel (c) to facilitate comparison (assuming the narrow H$\alpha$ emission is unchanged). The data is shown in black and the best fit model is overplotted in blue. The individual components -- narrow lines, broad line, and absorption -- are plotted in green, red, and orange, respectively. The reported FWHM and luminosity refer to the broad emission component shown in red. The uncertainties are dominated by systematics. \label{fig:sdss_spec}}
\end{figure*}

Gemini Director's Time observations of PHL 293B (GN-2019B-DD-109, P.I. Baldassare) were taken on 20 December 2019. Spectra were taken using the Gemini Multi Object Spectrograph (GMOS) on Gemini North. We used the 0.75$^{\prime\prime}$ slit with the R831\_G5302 grating, yielding a spectral resolution of $R\approx4500$. The central wavelength was set to $6600$~\AA, giving wavelength coverage from $5500$ -- $7500$~\AA. The seeing was 0.65$^{\prime\prime}$.

Spectra were reduced following the steps laid out in the GMOS Cookbook for the reduction of long-slit spectra with PyRAF \footnote{\url{http://ast.noao.edu/sites/default/files/GMOS\_Cookbook/}}. These include bias subtraction, flat-field correction, wavelength calibration, cosmic ray rejection, and flux calibration using the flux standard. The Gaussian spectral fitting is shown in Fig.~\ref{fig:sdss_spec}, and was done with the \textsc{PyQSOFit} code \citep{Guo2018,Shen2019}. We fit a continuum and Gaussian emission/absorption lines within user-defined windows and constraints on their widths.The continuum is modeled as a blue power-law plus a 3rd-order polynomial for reddening. The total model is a linear combination of the continuum and single or multiple Gaussians for the emission lines. Since uncertainties in the continuum model may induce subtle effects on measurements for weak emission lines, we first perform a global fit to the emission-line free region to better quantify the continuum.

We then fit multiple Gaussian models to the continuum-subtracted spectrum around the H$\alpha$ emission line region locally. For the SDSS H$\alpha$, we fit three Gaussians to model the narrow emission, broad emission, and absorption line. We use one narrow and one broad Gaussian to model the Gemini H$\alpha$ emission. Narrow Gaussians are defined as having $\rm{FWHM}<500$~km~s$^{-1}$. The narrow and broad line centroids are fit within a window of $\pm65$~\AA\ and $\pm100$~\AA, respectively. We use 100 Monte Carlo simulations to estimate the uncertainty in the line measurements. The spectral fitting of the H$\alpha$-[N~\textsc{ii}] complex is shown in Fig.~\ref{fig:sdss_spec} along with the SDSS spectral epoch taken in 2001 for comparison.

\section{Discussion}

\subsection{Comparison with Existing Observations} \label{sec:obs}

\subsubsection{Photometry}

The earliest photometry for PHL 293B is reported in \cite{Kinman1965} from the Palomar Sky Survey in 1965 and the Lick Observatory Carnegie Astrograph in 1949. \cite{Kinman1965} writes, ``a very rough estimate of the B magnitude is 17.7 on the Sky Survey plates and about a half-magnitude fainter on the 120-inch plates.'' \cite{Cairos2001} report a B magnitude of 17.67 in October 1988. The source is no brighter in the infrared than our comparison field star in digitized Sky Survey data taken on September 1995.

The galaxy's morphology was studied by \cite{Tarrab1987}. Later, the light profile was studied in more detail by \cite{Micheva2013}, who obtained deep multi-band imaging using the Nordic Optical Telescope in 2001. \cite{Micheva2013} report a B magnitude of 17.3. However, this difference can perhaps be attributed to instrument errors and different aperture definitions.

Catalina Sky Survey (CSS) photometry is available between April 2005 and October 2013 (roughly in-between the SDSS and DES photometry) and is analyzed in \cite{Terlevich2014}. However, the scatter in the CSS photometry is $\sim$ 0.1 \emph{V} mag. Still, we can rule out any large variability greater than one tenth of a magnitude in-between the SDSS and DES observations shown in Fig.~\ref{fig:1}. Our SDSS+DES light curve is consistent with \cite{Terlevich2014}, who constrain any variability over a few years to less than $0.02$ mag and over 25 years to less than a few tenths of magnitude. There is a gap in the photometry between September 1995 and the beginning of the SDSS data in September 1998.

The lack of large photometric variability and lack of spectral variability seen in PHL 293B prior to this work resulted in \cite{Terlevich2014} and \citet{Tenorio-Tagle2015} concluding that the source is not transient in nature. Given the small-amplitude optical variability from our precise DIA photometry and the recent dissipation of the unusually persistent broad H$\alpha$ emission, it is now clear that this in not the case. A complete re-interpretation of PHL 293B is now warranted.

\subsubsection{Spectroscopy}

\begin{figure*}
\gridline{\fig{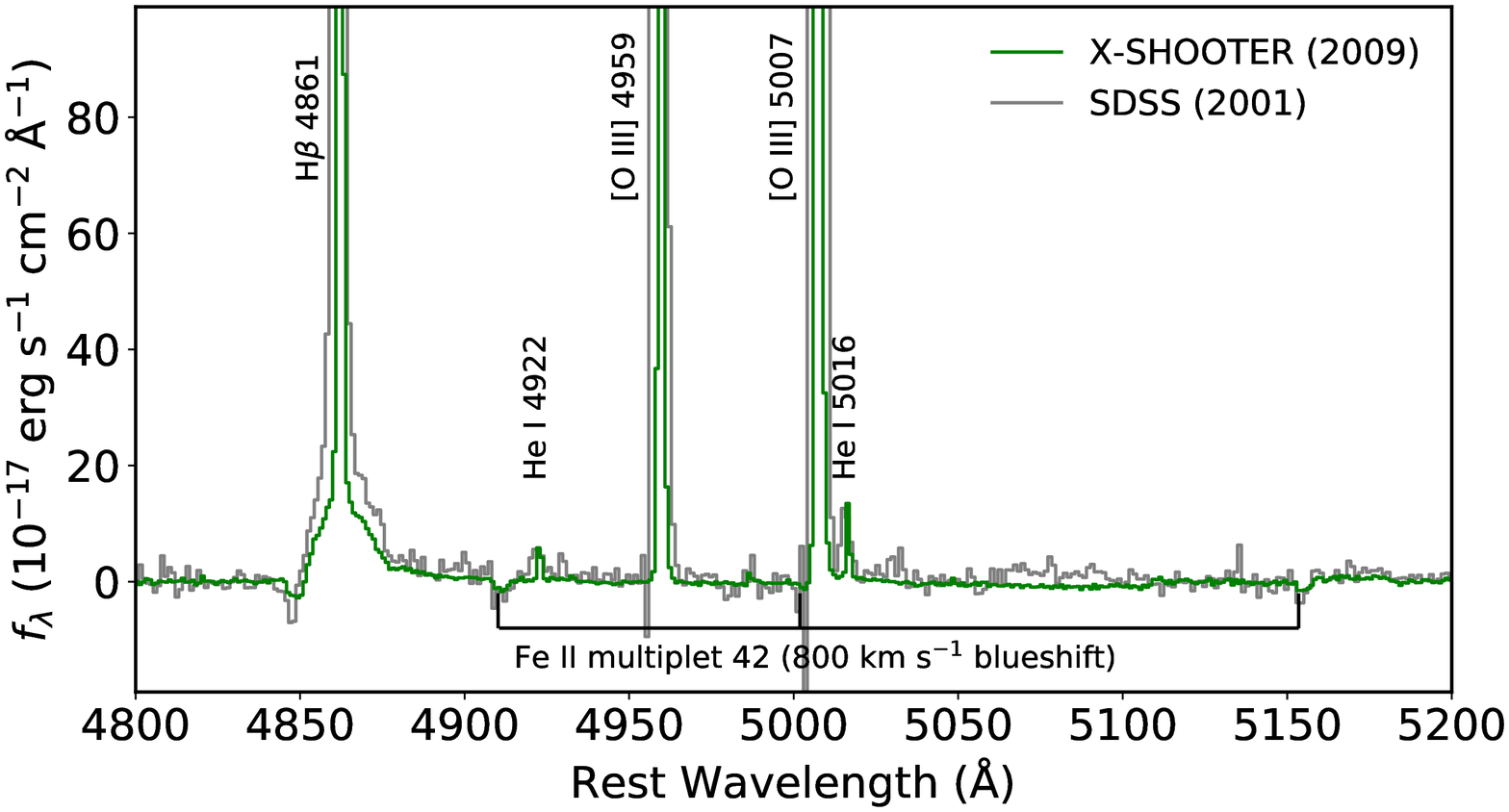}{0.5\textwidth}{(a)}
          \fig{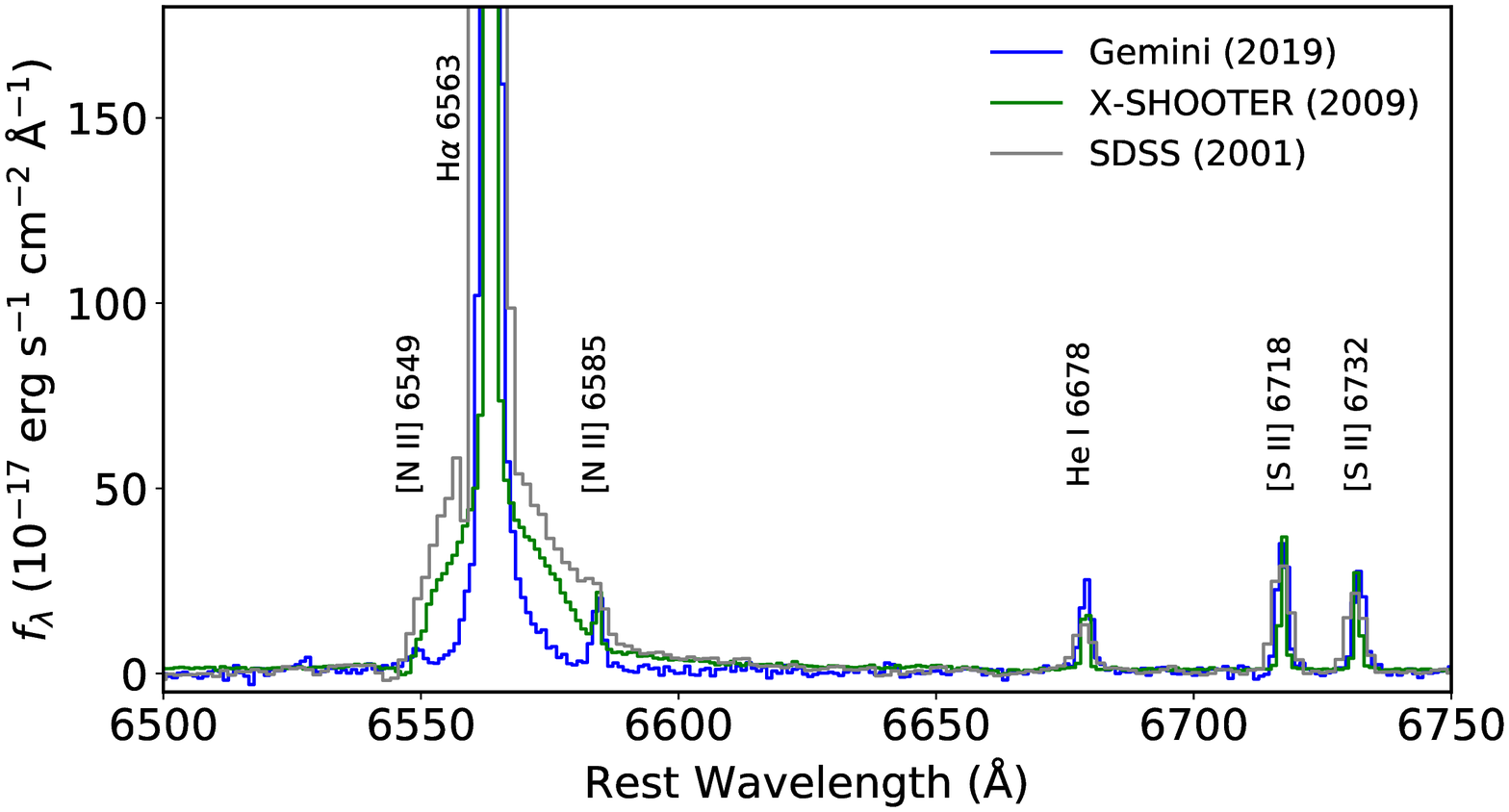}{0.5\textwidth}{(b)}
          }
\caption{Comparison of the SDSS, X-SHOOTER \citep{Izotov2011}, and Gemini GMOS-N spectra near H$\beta$ (a) and H$\alpha$ (b). All spectra are binned to $\Delta\log{\lambda}=10^{-4}$, flux-calibrated, de-reddened, redshift-corrected, and continuum model-subtracted. Differences between the SDSS and X-SHOOTER spectra can be attributed to aperture effects. The P Cygni-like absorption features are clearly present in the X-SHOOTER (1$^{\prime\prime}$ slit width) and SDSS (3$^{\prime\prime}$ fiber) data. The Gemini spectrum clearly shows the broad emission has begun to fade and the absorption feature is gone. \label{fig:3}}
\end{figure*}

The earliest spectrum of PHL 293B is shown in \cite{Kinman1965}. The presence of any broad emission is difficult to distinguish from the noise considering the [O~\textsc{iii}] $\lambda$4363 line is barely detected in their spectrum. Later, \cite{French1980} studied the line fluxes and its chemical abundance.

In 2001, PHL 293B was observed spectroscopically with SDSS. Broad and narrow Balmer emission with P Cygni profiles are clearly present (Fig.~\ref{fig:3}). \cite{Izotov2009} study the galaxy with archival UVES spectroscopy on the Very Large Telescope taken in November 2002. These authors conclude the spectral features are due to an LBV, which is reiterated in \cite{Izotov2011} with UV/optical and near-infrared X-SHOOTER spectroscopy obtained in August 2009 (60.A-9442(A), P.I. Diaz). \cite{Terlevich2014} note the presence of Fe~\textsc{ii} multiplet 42 and infrared Ca~\textsc{ii} triplet absorption lines blueshifted by the same velocity as the P Cygni absorption. They also obtained ISIS spectroscopy with the William Herschel Telescope in November 2011 and note no significant change in the broad emission between any of the spectra. We show both the SDSS and X-SHOOTER spectra in Fig.~\ref{fig:3} for comparison with our newly obtained GMOS-N spectrum. Our analysis of the SDSS spectrum H$\alpha$-[N~\textsc{ii}] complex is shown in Fig.~\ref{fig:sdss_spec}.

We fit a blue power-law continuum of $f_\lambda\sim6.4\times10^{-17}$ ($4.0\times10^{-17}$) erg s$^{-1}$ cm$^{-2}$ \AA$^{-1}$ at 6500 \AA\ with index $-3.1$ ($-5.0$) and reddening of $f_\lambda\sim5.2\times10^{-17}$ ($2.8\times10^{-17}$) erg s$^{-1}$ cm$^{-2}$ \AA$^{-1}$ at 6500 \AA\ in the SDSS (Gemini) spectrum.

In the Gemini spectrum, the broad and narrow H$\alpha$ luminosity is $2.6\times10^{38}$ and $2.8\times10^{39}$ erg s$^{-1}$, respectively. In the earlier SDSS spectrum, the broad and narrow H$\alpha$ luminosity is $1.3\times10^{39}$ and $3.2\times10^{39}$ erg s$^{-1}$, respectively. That is, a broad to narrow H$\alpha$ ratio of 0.41 with SDSS and 0.10 today with Gemini. The ratio of broad to narrow H$\alpha$ full-width-at-half-maximum (FWHM) is 8.5 in the SDSS fitting versus 5.9 in the Gemini fitting. The narrow absorption component is blueshifted by $807\pm65$ km s$^{-1}$ relative to H$\alpha$ in our model. The broad emission component in the Gemini spectrum is redshifted by $88\pm65$ km s$^{-1}$ relative to H$\alpha$. A P Cygni-like absorption feature is clearly present in the earlier SDSS (2001) and X-SHOOTER (2009) spectra. The absorption feature is not clearly visible in Gemini data, therefore no absorption component was used in our Gemini spectral fitting. 

Given the seeing of 0.65$^{\prime\prime}$ and the 0.75$^{\prime\prime}$ slit width, a decrease of 17 percent is expected with respect to the same source observed with an SDSS fiber, assuming the emission is dominated by a Gaussian point source. We indeed measured a decrease in the narrow line flux of 14 percent compared to the larger 3$^{\prime\prime}$ SDSS fiber. Variations in the narrow lines, therefore, are due to instrument/aperture effects. We measured a decrease in the broad H$\alpha$ of 80 percent. If 14 percent can be attributed to systematics from the aperture differences, the broad H$\alpha$ still decreased by about 66 percent. In addition, both the X-SHOOTER and ISIS spectra used a slit width of 1$^{\prime\prime}$ and the broad emission features were found to be unchanged with SDSS \citep{Izotov2011,Terlevich2014}. An 80 percent decrease could be explained if the broad emission is offset by 0.61$^{\prime\prime}$ from the slit center, inconsistent with Fig.~\ref{fig:1} (c) which shows the variable source is well-covered by the slit configuration.

Our Gemini spectrum does not cover the Fe~\textsc{ii} absorption lines, therefore we cannot determine if the Fe~\textsc{ii} absorption has weakened or disappeared. Recently, \citet{Allan2020} also report the fading of the broad Balmer emission and disappearance of the P Cygni absorption using new X-SHOOTER spectroscopy taken in December 2019.

\cite{Terlevich2014} model the H$\alpha$ emission with two broad emission components (one central and one redshifted extremely broad wing) and one narrow blueshifted absorption component. The ultra-broad red wing is no longer present in the Gemini data. A fading ultra-broad red wing was also seen in SDSS1133 \citep{Koss2014}.

Inspection of archival \emph{HST} Cosmic Origins Spectrograph far-UV spectrum shows C~\textsc{iii} $\lambda$1909 and C~\textsc{iv} $\lambda$1549 lines clearly present along with geocoronal Ly-$\alpha$ and O~\textsc{i}/Si~\textsc{ii}.

\subsubsection{X-Ray}

PHL 293B was not detected in 2009 with a 7.7 ks \emph{Chandra} exposure. The upper-limit on the X-ray luminosity is $\sim2.2\times10^{38}$ erg s$^{-1}$. In the AGN scenario, this implies an Eddington ratio below $10^{-5}$ assuming $M_\bullet=10^5\rm\ M_\odot$.


\subsubsection{Radio}

There is no detection in NVSS (1993), FIRST (2002) or VLASS images. From the VLASS sensitivity, we derive an upper limit of on the 5 GHz radio luminosity of $\nu L_{\nu}\sim4\times10^{35}$ erg s$^{-1}$ (assuming a flat spectral index).


\subsection{Conflicting Interpretations \& Likely Scenarios} \label{sec:discussion}

To understand the nature of PHL 293B, we investigate the following possible scenarios: a) low-mass AGN driven by a massive black hole, b) young stellar wind, c) tidal disruption event, d) luminous blue variable (LBV) star outburst, or e) long-lived SN IIn observed at late-times. In Table~\ref{tab:1}, we summarize the major observational properties of PHL 293B and evaluate each scenario. Scenarios (a) and (b) are unlikely given the recent fading of broad H$\alpha$ emission.

The origin of the narrow emission lines is likely the H~\textsc{ii} region ionized primarily by stellar emission from the massive star cluster. The observed continuum variability of $\sim$10 percent implies that most of the continuum originates from the cluster and nebular region. The presence of the high-velocity Fe~\textsc{ii} absorption implies a relatively cool medium in front of the continuum source, both extended by tens of parsecs. Therefore the Fe~\textsc{ii} absorption should also be extended by at least several tens of parsecs, implying a shell or LBV wind expanding at $\sim$800 km s$^{-1}$.

The fading of the light curve and broad emission is unusual for non-transient phenomena such as a stellar wind. This challenges the interpretation of \cite{Terlevich2014} of a superwind driven by a young stellar wind. Also, \cite{Tenorio-Tagle2015} point out that the dynamical time of an expanding shell with speed 800 km s$^{-1}$ would put the shock well outside of the galaxy given the age of the star cluster. Our observations also warrant a more skeptical look at the strongly-radiative stationary cooling wind, possibly driven by old supernova remnants, as proposed by \cite{Tenorio-Tagle2015}. While many observational features of PHL 293B can be explained by these scenarios, these authors assume no strong photometric or spectral variability in their models.

A tidal disruption event would also be highly unusual given the P Cygni absorption and unusually long-lived broad emission and small-amplitude photometric variability. Furthermore, the broad line width of $\sim 1500$ km s $^{-1}$ is several times smaller than expected for a typical tidal disruption event \citep{Arcavi2014}.

The optical and spectroscopic variability could arise from two different mechanisms (e.g. the optical variability could be purely stellar in origin with the P Cygni-like feature arising from another source). However, in this Letter we restrict ourselves to the simplest single mechanism which most likely explains all the observed features. Therefore, we argue the nature of the spectral features and variability of PHL 293B is due to a long-lived transient event. We devote the remainder of this section to investigating the remaining likely scenarios of an LBV outburst or long-lived SN IIn.

\subsubsection{Luminous Blue Variable Star Outburst}

LBVs are massive stars in a critical phase in stellar evolution located in the upper-left of the H-R diagram \citep{Humphreys1994}. Every star more massive than about $50\rm\ M_\odot$ will go through the LBV phase. Therefore, LBVs should be more common in BCD galaxies because the high star formation rates and low metallicities enables formation of massive stars.

\cite{Izotov2009} and \cite{Izotov2011} argue that the P Cygni-like spectral features of PHL 293B are evidence for an LBV in its star-forming region. Many of the spectral features observed in PHL 293B such as broad emission, narrow blueshifted absorption lines, and Fe~\textsc{ii} multiplet and Ca~\textsc{ii} infrared triplet lines are seen in LBV spectra \citep{Munari2009,Humphreys2017}. These features can often be seen in the long and fainter quiescent S Doradus phases. However, this seems inconsistent with the high luminosity of the LBV of $2.5$ -- $5.0\times10^6\ \rm{L_\odot}$ found by \citet{Allan2020}.

In addition, \cite{Terlevich2014} argue that the properties of PHL 293B are unusual of observed LBVs. The 800 km s$^{-1}$ blueshifted terminal velocity would be the largest ever reported for an LBV (with $\eta$ Carinae at $\sim 500$ km s$^{-1}$; \citealt{Leitherer1994}). These differences are attributed to the LBV undergoing a strong outburst and to effects at very low metallicity by \cite{Izotov2009}. However, this is not consistent with the \emph{lower} blueshifted absorption velocities of LBVs in the two low-metallicity dwarf galaxies NGC 2366 ($\sim 250$ km s$^{-1}$; \citealt{Drissen1997}) and IC 1613 ($\sim 300$ km s$^{-1}$; \citealt{Herrero2010}).

\cite{Izotov2011} also note broad H$\alpha$ luminosity of PHL 293B of $\sim10^{39}$ erg s$^{-1}$ from 2001--2009 is about ten times greater than the LBV in NGC 2366 \citep{Petit2006} when corrected for extinction. They conclude the LBV in PHL 293B must therefore be amongst the most luminous known and undergoing a strong outburst. However, the lack of any large photometric variability in any of the photometric epochs dating back to 1949 is challenging to this scenario.

LBVs, even in extragalactic settings, typically exhibit baseline variability of roughly 1 mag or larger on decade timescales, with sporadic and unpredictable eruptions observed in many cases \citep{Walborn2017}. In particular, P Cygni and $\eta$ Carinae have undergone massive outbursts of several magnitudes in recorded history. Although a short outburst may have been missed by the photometric gaps in PHL 293B, LBVs should continue to exhibit both photometric variability greater than $0.1$ mag even after an outburst \citep[e.g.][]{Smith2011}. Although long-term trends at this level have been observed in some LBVs, there is no evidence of the expected variability on shorter timescales in PHL 293B. In addition, demonstrable variability of spectral features during or shortly after an eruption should be observed \citep[e.g.][]{Richardson2011,Petit2006}. Any LBV in a relatively quiescent state that might explain the low-level of photometric variability in PHL 293B is difficult to reconcile with the long-lived broad emission lines and high terminal speed of the shock.

\citet{Allan2020} conclude the LBV is exiting its eruptive phase, is becoming obscured, or has collapsed directly into a black hole without producing a bright SN. The authors do not comment on Fe~\textsc{ii} absorption line variability. We expect these absorption features to be gone too if the shell of ejecta has expanded considerably or if the LBV has disappeared or become obscured.

\subsubsection{Long-Lived Type IIn Supernova} \label{sec:sn}

\begin{figure*}
\centering
\includegraphics[width=0.85\textwidth]{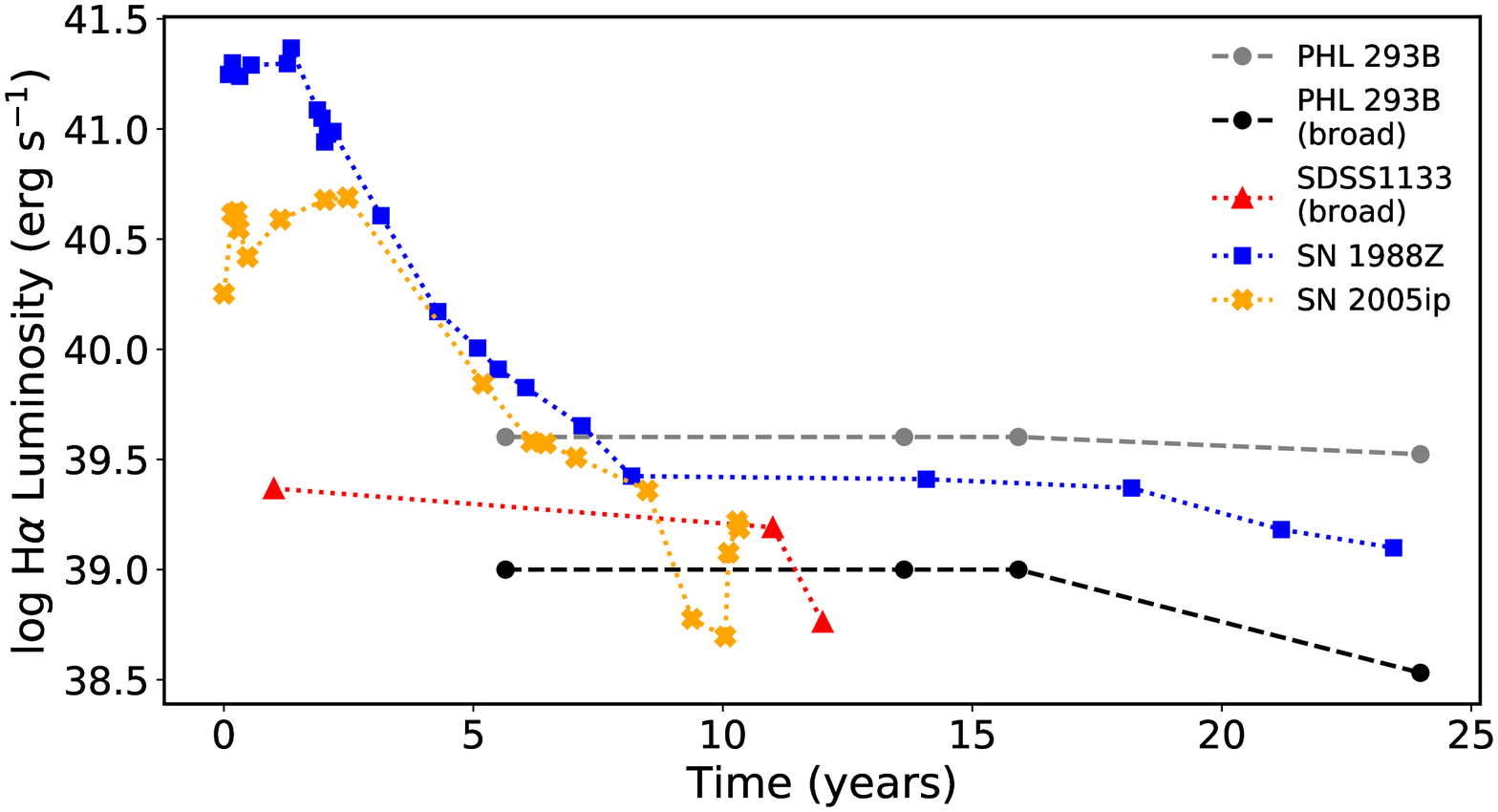}
\caption{H$\alpha$ luminosity versus time. SDSS1133 \citep{Koss2014}, SN 1988Z \citep{Aretxaga1999,Smith2017}, and SN 2005ip \citep{smith2009,Smith2017} are shown for comparison. The measurement refers to the total H$\alpha$ luminosity if not specified. We assume the SN in PHL 293B took place on January 1 1996. The data for PHL 293B prior to this work are estimated as $10^{39}$ erg s$^{-1}$ \citep{Izotov2009,Izotov2011,Terlevich2014} (whose measurements vary depending on how the broad emission is modeled, but are all consistent with $10^{39}$ erg s$^{-1}$). The UVES measurement was taken in non-photometric conditions and is therefore omitted. Uncertainties are dominated by systematic uncertainties, which are difficult to quantify exactly, but are typically $\pm$20 percent. \label{fig:haevol}}
\end{figure*}

If the features of PHL 293B are due to an evolving young SN remnant, it falls into the rare Type IIn class. That is, a core-collapse SN with broad and narrow Balmer lines \citep{Schlegel1990,Filippenko1997}. P Cygni-like narrow absorption lines are commonly observed in SNe IIn \citep[e.g.][]{Salamanca2002}. The spectral features are due to the ejecta from a massive progenitor star interacting with the circumstellar medium. There is evidence that this scenario arises when a strong mass loss from an LBV-like progenitor undergoes sustained interaction with a dense circumstellar medium (e.g. \citealt{Koss2014,Smith2017} and references within).

Some Type I and II SNe show high-velocity blueshifted Fe~\textsc{ii} lines \cite[e.g.][]{Chugai2004,Young2010,Smith2010} and infrared Ca~\textsc{ii} triplet lines \cite[e.g.][]{Taddia2013}. Missing associated Fe~\textsc{ii}/Ca~\textsc{ii} emission lines could be due to an asymmetric explosion, or the emission lines could be very weak and broad at late-times making them difficult to detect. The 800 km s$^{-1}$ blueshift of the absorption features in PHL 293B's spectrum is not unusual for an expanding envelope of SN IIn ejecta which can exceed 1000 km s$^{-1}$ in some cases \citep[e.g. SN 2009ip;][]{Foley2011,Smith2014}, but is anomalously high for an LBV wind. Furthermore, the broad emission width in the SDSS and X-SHOOTER spectra of FWHM $\sim 10^3$ km s$^{-1}$ is not unusual \citep[e.g.][]{Nyholm2017}. Sustaining broad emission of $10^{39}$ erg s$^{-1}$ over 15 years requires an energy of $\sim 5\times10^{47}$ ergs, well below the canonical SN kinetic energy budget of $10^{51}$ ergs.

A remarkably similar transient known as SDSS1133 was discovered by \cite{Koss2014} in the BCD galaxy Mrk 177. Like PHL 293B, its broad emission persists for over a decade with broad $L_{\rm{H}\alpha}\sim7\times10^{39}$ erg s$^{-1}$ at day 4000. SDSS1133 also showed an ultra-broad H$\alpha$ emission. A growing number of \emph{bona fide} SNe IIn are known to display broad emission that persists for a decade or longer. We show the broad H$\alpha$ luminosity versus time in Fig.~\ref{fig:haevol} in comparison with the long-lived SNe IIn SDSS1133, SN 1988Z, and SN 2005ip \citep{Aretxaga1999}. The SDSS spectra of SDSS1133 and PHL 293B are shown for comparison in Appx.~\ref{apdx2}.

It is possible the SN IIn-like event occurred sometime between September 1995 and September 1998 when no photometry is available. In some cases, SNe IIn light curves exhibit ``bumps'' several years after the explosion \citep[e.g.][]{Nyholm2017}.

The VLASS upper limit implies the transient is not radio-loud like some very luminous SNe \citep{Smith2017}. In the case of SDSS1133, \emph{Swift} detected X-ray emission with an estimated X-ray luminosity of $1.5\times10^{39}$ erg s$^{-1}$ 12 years after the SN \citep{Koss2014}. The X-ray upper limit of PHL 293B of $2.2\times10^{38}$ erg s$^{-1}$ in 2009 is an order of magnitude less than SDSS1133. Assuming the SN in PHL 293B took place in 1996, it was observed with \emph{Chandra} $\sim$13 years after its outburst. SN IIn generally emit X-rays above a few $10^{38}$ erg s$^{-1}$ even several years after the SN~\citep[e.g.][]{Bregman1992}, however their X-ray emission is very diverse and may approach the X-ray limit of PHL 293B 13 years after the outburst \citep[see Fig. 3 of ][]{Dwarkadas2012}. This perhaps indicates the outburst in PHL 293B was non-terminal or the X-ray emission has declined steeply at late-times.

\begin{deluxetable*}{l|ccccc}
\tablenum{1}
\tablecaption{Source properties and possible scenarios}
\tablewidth{0pt}
\tablehead{
Source Property & \colhead{LBV outburst} & \colhead{Stellar Wind} & \colhead{AGN} & \colhead{SN IIn} & \colhead{TDE}
}
\decimalcolnumbers
\startdata
Broad emission lines ($L_{\rm H\alpha}\sim10^{39}$ erg s$^{-1}$)    & unusually large & YES & YES & YES & unusually low \\
Decade long-lived broad emission          & unusual & YES & YES & unusual & NO \\
Recent dissipation of broad emission               & YES & NO & NO & YES & YES \\
P Cygni-like profile                               & YES & YES & unusual & YES & NO \\
Fe~\textsc{ii} absorption lines                    & YES & YES & unusual & YES & NO \\
800 km s$^{-1}$ blueshift of absorption lines       & unusually high & NO & YES & YES & NO \\
Lack of X-rays ($\lesssim 2.2\times10^{38}$ erg s$^{-1}$) & YES & YES & NO & unusual & YES \\
Lack of high ionization lines ($\lesssim 120$ eV)   & YES & YES & unusual & YES & YES \\
small-amplitude optical variability & unusually small & NO & YES & YES & unusually low \\
\enddata
\tablecomments{YES: the observational feature can be readily explained by the proposed scenario. NO: cannot be explained without invoking an exotic or contrived scenario. Unusual: can be explained but it is unusual of observed systems.
\label{tab:1}}
\end{deluxetable*}

\section{Conclusions} \label{sec:conclusion}

The most plausible explanations for the recent dissipation of the broad emission after an unusually persistent phase are an LBV outburst followed-by a slow, weakly variable phase or a very long-lived SN IIn event. The latter is more likely given the lack of short-timescale variability and the slowly-fading light curve. The similarity to the persistent transient of \cite{Koss2014} in a similar BCD galaxy is of note. In this case, the SN occurred sometime between 1988 and 1998 and continued to slowly evolve until today. However, in our case, the evidence for an LBV progenitor is only circumstantial.

The question, ``why are there not more dwarf starburst galaxies with broad emission as strong as PHL 293B?'' is raised when considering the stationary wind scenario. The unusual nature of PHL 293B and SDSS1133 can be well-understood if they are due to rare stellar transient phenomena. However, high-resolution spectroscopic observations should be conducted in coming years to study its spectral behavior in detail.

We reiterate the warning of \cite{Filippenko1989} that the long-lived spectral features seen of some SNe II can be AGN impostors. This is the case especially at late-times when the features are due to ejecta-circumstellar medium interaction as well as SNe at low bolometric luminosities. We extend this warning to low-mass AGN, noting that late-time SN variability can mimic AGN-like variability~\citep[also see][]{Aretxaga1997,Aretxaga1999b}. Perhaps the dense circumstellar medium of metal-poor massive stars plays an important role in extending the lifetime of the broad emission. \cite{Izotov2008} and \cite{Izotov2010} identified a handful of low-metallicity compact emission-line galaxies with persistent broad H$\alpha$ emission. They suggested these systems could be low-mass AGNs. However, no characteristic hard X-ray emission was seen in \emph{Chandra} images \citep{Simmonds2016, 2017ApJ...836...20B}. If the sample of \cite{Izotov2008}, \cite{Izotov2009sn}, \cite{Izotov2010}, \cite{Koss2014}, and PHL 293B can be explained by similar mechanisms, then perhaps such long-lived transients are more common in low-metallicity dwarf galaxies. However, the H$\alpha$ luminosity of the \cite{Izotov2008} sample is much larger than PHL 293B. In these scenarios, only follow-up spectroscopy on decade-long or greater timescales can be definitive.

Analysis of difference imaging light curves with the Legacy Survey of Space and Time at Vera C. Rubin Observatory would determine if the source continues to exhibit low-levels of variability or not. We expect little or no continued variability if the SN was terminal. However, we expect continued small-amplitude variability if the progenitor is still present.

\acknowledgments

We thank Yuri Izotov for helpful correspondence. We thank Tom Diehl and Chris Lidman for helpful comments. C.J.B. is grateful to Kedar Phadke and Gautham Narayan for useful discussion, and to the Illinois Graduate Survey Science Fellowship for support. We thank the referee, Roberto Terlevich, for useful comments and corrections which improved this work.

This research has made use of the NASA/IPAC Extragalactic Database (NED), which is operated by the Jet Propulsion Laboratory, California Institute of Technology, under contract with the National Aeronautics and Space Administration.

Based on observations obtained at the Gemini Observatory, which is operated by the Association of Universities for Research in Astronomy, Inc., under a cooperative agreement with the NSF on behalf of the Gemini partnership: the National Science Foundation (United States), National Research Council (Canada), CONICYT (Chile), Ministerio de Ciencia, Tecnolog\'{i}a e Innovaci\'{o}n Productiva (Argentina), Minist\'{e}rio da Ci\^{e}ncia, Tecnologia e Inova\c{c}\~{a}o (Brazil), and Korea Astronomy and Space Science Institute (Republic of Korea).

Support for V.F.B. was provided by the National Aeronautics and Space Administration through Einstein Postdoctoral Fellowship Award Number PF7-180161 issued by the Chandra X-ray Observatory Center, which is operated by the Smithsonian Astrophysical Observatory for and on behalf of the National Aeronautics Space Administration under contract NAS8-03060.

Funding for the Sloan Digital Sky Survey IV has been provided by the Alfred P. Sloan Foundation, the U.S. Department of Energy Office of Science, and the Participating Institutions. SDSS-IV acknowledges support and resources from the Center for High-Performance Computing at the University of Utah. The SDSS web site is www.sdss.org.

SDSS-IV is managed by the Astrophysical Research Consortium for the 
Participating Institutions of the SDSS Collaboration including the 
Brazilian Participation Group, the Carnegie Institution for Science, 
Carnegie Mellon University, the Chilean Participation Group, the French Participation Group, Harvard-Smithsonian Center for Astrophysics, 
Instituto de Astrof\'isica de Canarias, The Johns Hopkins University, Kavli Institute for the Physics and Mathematics of the Universe (IPMU) / 
University of Tokyo, the Korean Participation Group, Lawrence Berkeley National Laboratory, 
Leibniz Institut f\"ur Astrophysik Potsdam (AIP),  
Max-Planck-Institut f\"ur Astronomie (MPIA Heidelberg), 
Max-Planck-Institut f\"ur Astrophysik (MPA Garching), 
Max-Planck-Institut f\"ur Extraterrestrische Physik (MPE), 
National Astronomical Observatories of China, New Mexico State University, 
New York University, University of Notre Dame, 
Observat\'ario Nacional / MCTI, The Ohio State University, 
Pennsylvania State University, Shanghai Astronomical Observatory, 
United Kingdom Participation Group,
Universidad Nacional Aut\'onoma de M\'exico, University of Arizona, 
University of Colorado Boulder, University of Oxford, University of Portsmouth, 
University of Utah, University of Virginia, University of Washington, University of Wisconsin, 
Vanderbilt University, and Yale University.

Funding for the DES Projects has been provided by the U.S. Department of Energy, the U.S. National Science Foundation, the Ministry of Science and Education of Spain, 
the Science and Technology Facilities Council of the United Kingdom, the Higher Education Funding Council for England, the National Center for Supercomputing 
Applications at the University of Illinois at Urbana-Champaign, the Kavli Institute of Cosmological Physics at the University of Chicago, 
the Center for Cosmology and Astro-Particle Physics at the Ohio State University,
the Mitchell Institute for Fundamental Physics and Astronomy at Texas A\&M University, Financiadora de Estudos e Projetos, 
Funda{\c c}{\~a}o Carlos Chagas Filho de Amparo {\`a} Pesquisa do Estado do Rio de Janeiro, Conselho Nacional de Desenvolvimento Cient{\'i}fico e Tecnol{\'o}gico and 
the Minist{\'e}rio da Ci{\^e}ncia, Tecnologia e Inova{\c c}{\~a}o, the Deutsche Forschungsgemeinschaft and the Collaborating Institutions in the Dark Energy Survey. 

The Collaborating Institutions are Argonne National Laboratory, the University of California at Santa Cruz, the University of Cambridge, Centro de Investigaciones Energ{\'e}ticas, 
Medioambientales y Tecnol{\'o}gicas-Madrid, the University of Chicago, University College London, the DES-Brazil Consortium, the University of Edinburgh, 
the Eidgen{\"o}ssische Technische Hochschule (ETH) Z{\"u}rich, 
Fermi National Accelerator Laboratory, the University of Illinois at Urbana-Champaign, the Institut de Ci{\`e}ncies de l'Espai (IEEC/CSIC), 
the Institut de F{\'i}sica d'Altes Energies, Lawrence Berkeley National Laboratory, the Ludwig-Maximilians Universit{\"a}t M{\"u}nchen and the associated Excellence Cluster Universe, 
the University of Michigan, the National Optical Astronomy Observatory, the University of Nottingham, The Ohio State University, the University of Pennsylvania, the University of Portsmouth, 
SLAC National Accelerator Laboratory, Stanford University, the University of Sussex, Texas A\&M University, and the OzDES Membership Consortium.

Based in part on observations at Cerro Tololo Inter-American Observatory, National Optical Astronomy Observatory, which is operated by the Association of 
Universities for Research in Astronomy (AURA) under a cooperative agreement with the National Science Foundation.

The DES data management system is supported by the National Science Foundation under Grant Numbers AST-1138766 and AST-1536171.
The DES participants from Spanish institutions are partially supported by MINECO under grants AYA2015-71825, ESP2015-66861, FPA2015-68048, SEV-2016-0588, SEV-2016-0597, and MDM-2015-0509, 
some of which include ERDF funds from the European Union. IFAE is partially funded by the CERCA program of the Generalitat de Catalunya.
Research leading to these results has received funding from the European Research
Council under the European Union's Seventh Framework Program (FP7/2007-2013) including ERC grant agreements 240672, 291329, and 306478.
We  acknowledge support from the Brazilian Instituto Nacional de Ci\^encia
e Tecnologia (INCT) e-Universe (CNPq grant 465376/2014-2).

This manuscript has been authored by Fermi Research Alliance, LLC under Contract No. DE-AC02-07CH11359 with the U.S. Department of Energy, Office of Science, Office of High Energy Physics.

%

\vspace{5mm}
\facilities{SDSS, DES, Gemini(GMOS-N)}


\software{astropy, hotpants, PyQSOFit
          }



\appendix

\section{Light Curve Construction and Analysis} \label{apdx}

Difference imaging was performed independently on SDSS and DES data. The SDSS pipeline described in \cite{Baldassare2018} makes use of a modified version of the Difference Imaging and Analysis Pipeline 2 \citep{2000AcA....50..421W}. The DES difference pipeline is described in \cite{Kessler2015} and is based on \textsc{hotpants} \citep{Becker2013}. Both codes rely on the basic image subtraction algorithm described in \cite{1998ApJ...503..325A, 2000A&AS..144..363A}. Our procedure for constructing our DIA light curves is as follows: We construct a template image using single-epoch frames with the best seeing and lowest background. Next, we convolve each image with a kernel function to approximately match the PSF in the DES template image. Then we subtract the convolved single-epoch image from the template to create the difference image. Finally, we perform aperture photometry on the difference image within a 2.5$^{\prime\prime}$ radius circle centered on the target galaxy nucleus.

We invoke the \textsc{carma\_pack} software \citep{Kelly2018} to perform a Markov chain Monte Carlo fitting to a DRW model. In the framework of \cite{Kelly2014}, the standard DRW model is the continuous-time first-order autoregressive (${\rm CAR}(1)$) Gaussian process. To asses the fit, we calculate $\left[\chi^2/\nu\right]_{\rm DRW}$ from the standardized residuals of the DRW/${\rm CAR}(1)$ model, assuming the number of degrees of freedom is $N-2$ (for the 2 parameters in the ${\rm CAR}(1)$ process).

We also calculate the standard variability metric $\left[\chi^2/\nu\right]_{\rm var}$:
\begin{equation}
\left[\chi^2/\nu\right]_{\rm var} = \frac{1}{\nu}\sum^N_{i=1} (m_i-\overline{m})^2 w_i
\end{equation}
where the weighted mean $\overline{m}$ is given by,
\begin{equation}
\overline{m} = \frac{\sum^N_{i=1}m_i w_i}{\sum^N_{i=1} w_i}
\end{equation}
with weights given by the reciprocal of the photometric errors $w_i=1/\sigma_i^2$ on each measurement $m_i$ (in magnitudes). We then calculate the resulting significance $\sigma_{\rm var}$ from the $\chi^2$ distribution in units of $\sigma$. We find significant variability with $\sigma_{\rm var}>3$ in both SDSS and DES \emph{g} and \emph{r} band light curves for PHL 293B before combining the photometry.

\section{Comparison of Spectra to SDSS1133} \label{apdx2}

\begin{figure*}[ht]
\centering
\includegraphics[width=0.85\textwidth]{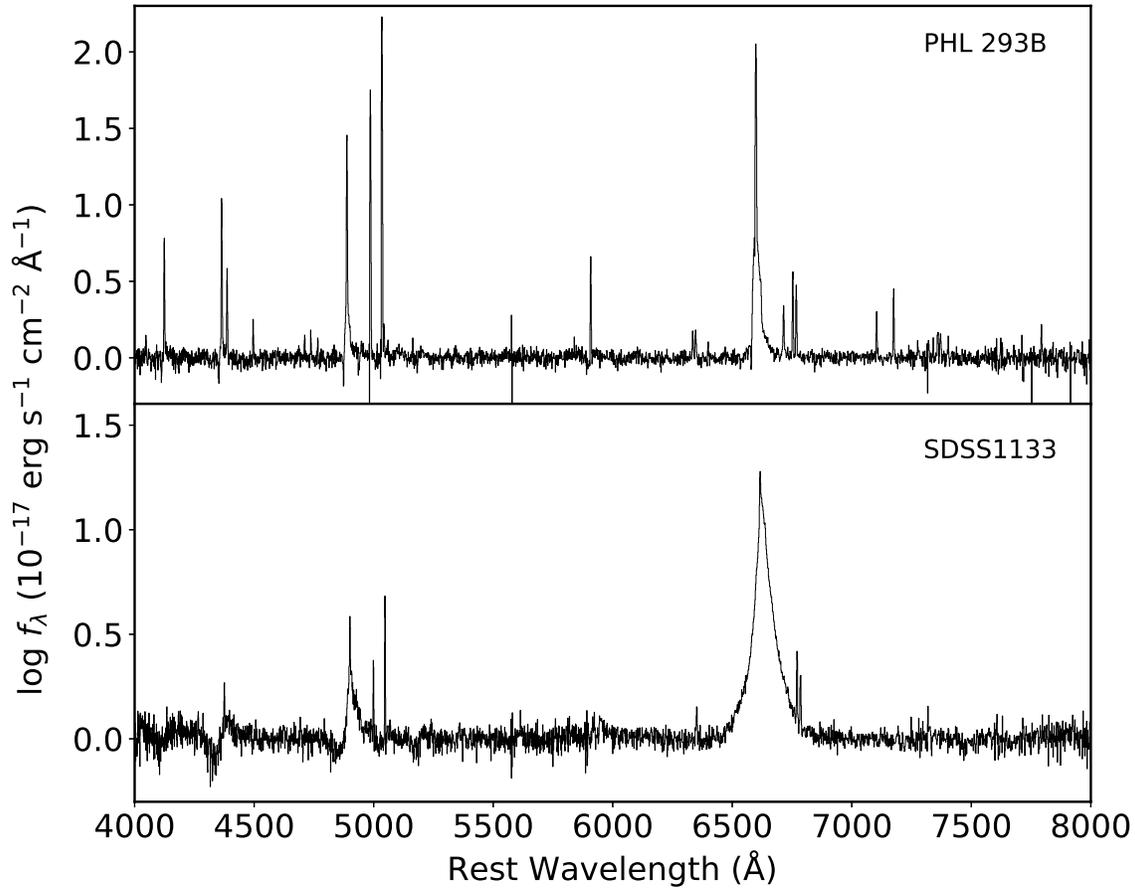}
\caption{Comparison of SDSS spectra of PHL 293B and SDSS1133.}
\end{figure*}


\bibliography{BibTex, sample63}
\bibliographystyle{aasjournal}

\allauthors


\end{document}